\pdfoutput=1
\documentclass[a4paper, amsfonts, amssymb, amsmath, reprint, showkeys, nofootinbib, twoside, superscriptaddress]{revtex4-2} 
\usepackage[english]{babel}
\usepackage[utf8]{inputenc}
\usepackage{enumerate}
\usepackage{lipsum}
\usepackage{braket}
\usepackage{csquotes}
\usepackage{hyperref}
\hypersetup{pdftitle={Article}, pdfauthor={Author}, colorlinks=false}
\usepackage[most]{tcolorbox}
\usepackage{booktabs} 
\usepackage{multirow} 
\begin{document}
\title{RotorMap and Quantum Fingerprints of DNA Sequences \\ via Rotary Position Embeddings}

\author{Danylo Yakymenko}
\email{danylo.yakymenko@gmail.com}
\affiliation{Institute of Mathematics of NAS of Ukraine, 3 Tereschenkivska str., Kyiv 01024, Ukraine.}
\affiliation{Kyiv Academic University, 36 Vernadsky blvd., Kyiv 03142, Ukraine.}
\author{Maksym Chernyshev}
\email{chernmakc@gmail.com }
\affiliation{Kyiv Academic University, 36 Vernadsky blvd., Kyiv 03142, Ukraine.}
\author{Illia Savchenko}
\email{illiasavchenkoo@gmail.com}
\affiliation{Institute of Physics of NAS of Ukraine, 46 Nauki avenu, Kyiv 03028, Ukraine.}
\author{Sergii Strelchuk}
\email{sergii.strelchuk@cs.ox.ac.uk}
\affiliation{Department of Computer Science, University of Oxford, Oxford, OX1 3QG, United Kingdom.}

\date{\today}

\begin{abstract}
For strings of letters from a small alphabet, such as DNA sequences, we present a quantum encoding that empirically provides a strong correlation between the Levenshtein edit distance and the fidelity between quantum states defined by the encodings. It is based on the principles of Rotary Position Embeddings (RoPE), employed in modern large language models. 
Classically, this encoding yields RotorMap — a GPU-accelerated DNA mapping algorithm that achieves speedups of 50–700$\times$ over single-thread Minimap2 in proof-of-concept tests on human and maize genomes.


For use on quantum devices, we introduce the Angular encoding, which is built from RoPE and directly outputs state preparation circuits. To verify its properties and utility on NISQ devices, we report results of experiments conducted on quantum computers from Quantinuum: the 56-qubit H2-1, H2-2 and the latest 98-qubit Helios-1. 
As a potential application, we consider a quantum DNA authentication problem and conjecture that a quantum advantage in one-way communication complexity could be achieved over any comparable classical solution.
\end{abstract}

\keywords{Levenshtein distance, approximate string matching, quantum fingerprint, rotary positional embedding, DNA mapping and alignment, quantum computing, angular encoding, RotorMap, quantum DNA authentication, gap edit distance, communication complexity}

\maketitle

\section{Introduction and results} \label{sec:introduction}

The classical fingerprint of a string is a shorter string associated with it that allows one to perform identification, communication, comparison, and related tasks in an efficient manner. Depending on the task conditions, various schemes can be regarded as fingerprinting, such as cryptographic hashes, lossy compressed representations, database indexes, etc. In modern AI systems, vector embeddings of texts (and other data) serve a similar purpose to fingerprints, as they enable efficient information retrieval and generation.

Quantum computing provides an entirely new instrument, as we can now use quantum states as fingerprints. To understand the basic principle, consider the following scheme: for a string $S = S_1S_2...S_n$ of length $n$, we can define the quantum fingerprint of $S$ by
$$\ket{Q(S)} = \frac{1}{\sqrt{n}} \sum_{i} \ket{i}\ket{S_i}$$
where we need $\log_2 n$ qubits in the first register to encode the position $\ket{i}$ of a letter, and $\log_2 L$ qubits in the second register to encode the letter from an alphabet of size $L$. For the fidelity between fingerprints, we then have
$$|\braket{Q(S)|Q(T)}|^2 = \left(\frac{n-d}{n}\right)^2$$
where $d$ is the Hamming distance (the number of mismatched letters at the same positions) between strings $S$ and $T$.
Fidelity between quantum states is a natural measure of their closeness. It equals the probability of misidentifying one state for another in quantum algorithms. This fingerprinting scheme thus provides a correlation between the Hamming distance in the space of strings and quantum state similarity in the Hilbert space, modeling a quantum system on a logarithmic number of qubits. Its combination with error-correcting codes was used in \cite{Buhrman2001Quantum} to prove an exponential advantage in quantum communication complexity over the classical setting for a specific communication problem.

In bioinformatics, however, the Hamming distance is of less interest, since deletions and insertions of base pairs are common in DNA and other coding sequences of biological organisms, either due to reading errors or mutations. Edit distances between genomic strings are introduced to formalize a more natural notion of similarity. An example of such an edit distance is the Levenshtein distance (LD) \cite{Levenshtein1966Binary}, which is equal to the minimal number of insertions, deletions, and substitutions required to transform one string into another. While the Hamming distance can be computed in linear time, algorithms for exact LD computation require nearly quadratic time. It is proved in \cite{Backurs2015Edit} that subquadratic time is unreachable if the Strong Exponential Time Hypothesis \cite{Impagliazzo2001} (SETH) is true. 
Despite this, the LD approximation can be computed more quickly and usually suffices in practice. For example, a near-linear time algorithm that approximates LD was constructed in \cite{Andoni2010Polylog}. See \cite{Berger2021Levenshtein, Koerkamp2025Optimal} for a collection of classical results on this topic and \cite{Gibney2024} for quantum. 

The edit distance computation and its complexity are particularly relevant in DNA mapping and alignment problems, where, given a read (a small DNA fragment, also called $k$-mer where $k$ is its length), one asks to find its most similar matching locations in a reference DNA sequence. The human genome DNA is approximately 3.3 billion base pairs in size, so a straightforward algorithm that runs through the entire reference sequence and computes edit distances between the read and reference fragments would be too far from being practical. Instead, modern DNA aligners use the so-called seed-chain-extend paradigm, see \cite{Shaw2023} for its complexity analysis. In this paradigm, a set of very small DNA pieces, called anchors or seeds and usually taken as minimizers or syncmers, are marked in the reference sequence in an evenly spaced fashion. Then, by looking at the sequence of seeds in a read, the algorithm finds a matching chain of seeds in the reference, finishing with the extension phase that aligns strings in the gaps between seeds. Currently, Minimap2 \cite{Li2018Minimap2, Li2021NewStrategies} is the most popular state-of-the-art algorithm that uses this approach, although the Bowtie2 \cite{Langmead2009Ultrafast, Langmead2012FastGapped, Langmead2018} or BWA \cite{Li2009FastShort, Li2010FastLong, Li2013AligningMEM} algorithm could be more suitable in some cases. When mapping long reads with low error rates (less than 1\%), the mapquick algorithm \cite{Ekim2023Mapquik} achieves a significant speedup over Minimap2 by combining consecutive seeds together.

\begin{figure}[h!]
    \centering
    \includegraphics[width=0.8\linewidth]{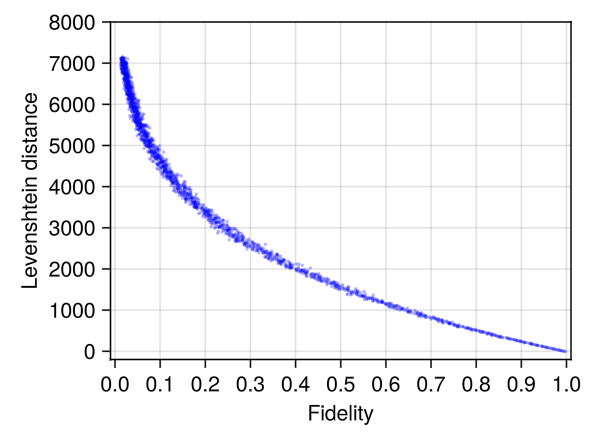}
    \caption{
For a random 20,000-long DNA sequence and its mutation, we compute the Levenshtein distance on the one hand and fidelity between their RoPE encodings on the other. In this case, a RoPE with 4096 complex dimensions is used, forming a quantum state on 12 qubits. In each pair, the introduced mutations are evenly split between random deletions, insertions and substitutions, where the mutation rate is selected evenly from the interval (0, 0.5). A total of n=2,000 sampled pairs are plotted. 
    }
    \label{fig:intro_corr}
\end{figure}

\begin{figure}[h!]
    \centering
    \includegraphics[width=1\linewidth]{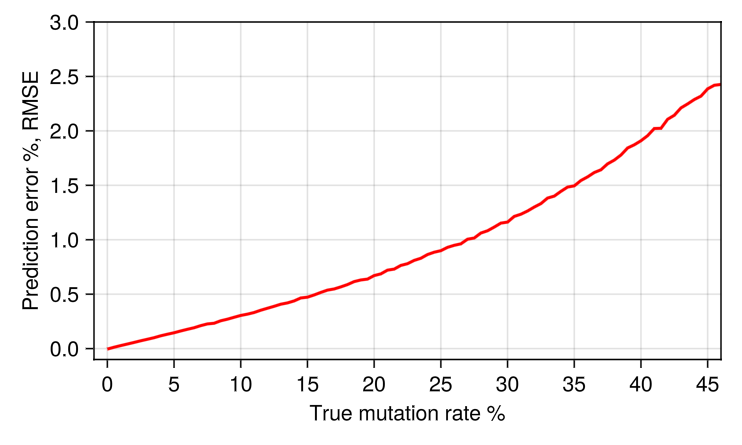}
    \caption{When sampling randomly a DNA sequence and its mutation, we can predict the mutation rate based on the observed fidelity between encodings. For the encoding shown in Fig. \ref{fig:intro_corr}, the RMSE is less than 1\% if the mutation rate is less than 25\%. 
    }
    \label{fig:intro_pred}
\end{figure}

\begin{figure}[h!]
    \centering
    \includegraphics[width=1\linewidth]{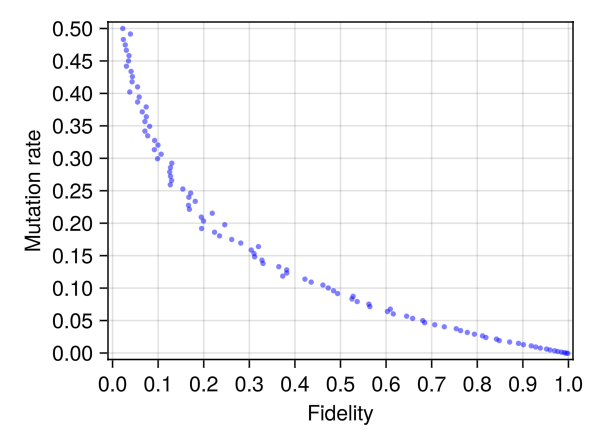}
    \caption{100 random DNA sequences of length 1 billion each were generated along with their mutations. Since it is infeasible to find the exact Levenshtein distance in this case, we plot the correlation between the mutation rate and the fidelity between respective RoPE encodings of dimension 2048.
    }
    \label{fig:intro_1bil}
\end{figure}

In this work, we present a novel way to encode DNA sequences as quantum states that respects the Levenshtein distance. For example, Figure \ref{fig:intro_corr} shows the obtained correlation between LD and fidelity for 20,000-long random DNA strings encoded as 12-qubit quantum states. By examining the fidelity, this correlation allows us to predict LD within only a few percent of the error, as shown in Figure \ref{fig:intro_pred}. Surprisingly, the encoding works even for super long DNA strings, which have an order of 1 billion letters, as seen in Figure \ref{fig:intro_1bil}. Non-random DNA strings with specific repeat patterns require some care, as we discuss later, but they do not pose a fundamental issue. Our DNA encoding is built using the principles of Rotary Positional Embeddings (RoPE) \cite{Su2024RoFormer, Barbero2025Round} utilized in modern Large Language Models (LLM), so we call it RoPE-based, RoPE-DNA or simply the RoPE encoding. The details of its construction can be found in Section \ref{sec:rope}.

Even though the RoPE-DNA encoding defines a quantum state, we can also use it classically as a high-dimensional complex vector. The classical complexity of computing this encoding vector is linear in the DNA string length, which allows us to estimate LD very quickly by computing the inner product between encodings. Moreover, if we want to estimate pairwise Levenshtein distances between two batches of strings, the problem essentially reduces to matrix multiplication. Fortunately, modern GPU and TPU hardware have accumulated decades of research to make matrix multiplication run as fast as possible. The RoPE-DNA computation also has a fixed predictable branching, thus allowing a full utilization of the SIMT (Single Instruction, Multiple Threads) execution model. Based on these observations, we have designed a novel, RoPE-based GPU-accelerated DNA mapping algorithm, called RotorMap, described in Section \ref{sec:rotormap}. Our proof-of-concept implementation of RotorMap is developed in the Julia programming language \cite{bezanson2017julia} using the CUDA.jl library \cite{besard2018juliagpu, besard2019prototyping}. In tests, the approximate mapping of a batch of about 96,000 DNA reads of length 20,000 bp each to the human genome (in both positive and negative directions) takes less than 40 seconds on a single H100 NVIDIA GPU, assuming preloaded data. In contrast, Minimap2 takes about 50 seconds to map the same data on 80 virtualized CPU threads of Intel(R) Xeon(R) Gold 6448Y and on the order of 2000 seconds if a single thread is used. When testing the mapping of about 64,000 reads to the maize genome, the running time of RotorMap is less than 20 seconds, while Minimap2 takes about 280s on 80 threads and 14,000s on a single thread. However, the results of the current version of RotorMap are not entirely perfect and may require additional time to refine. On the other hand, the development of RotorMap is still work in progress, thus we believe a genuine 1000-fold speedup over Minimap2, running on a single CPU thread, is achievable. More details of the tests can be found in Section \ref{sec:rotormap}.

Despite the apparent theoretical advantage of using the RoPE encoding as a quantum state (we only need a logarithmic number of qubits, after all), it is not trivial to utilize it in practice. The first issue arises from the fact that to work with the encoding on quantum computers, one has to prepare it using a state preparation quantum circuit. Finding such circuits that produce the desired states, even approximately, is a genuinely hard problem. One way of solving it is based on using matrix product states (MPS), see \cite{creevey2025scalablequantumstatepreparation} for example. In general, however, preparation circuits should contain an exponential number of primitive quantum gates, making it less viable for high-qubit encodings. To address this issue, we have designed another encoding, called Angular, which is built from RoPE. The Angular encoding of a DNA string produces a quantum preparation circuit directly. Even though the state prepared by this encoding does not actually match the RoPE encoding, the correlation between LD and fidelity is largely preserved, see Section \ref{sec:angular} for the details.

The Angular encoding also helps us deal with the second obstruction in quantum computing, namely, the quantum noise in currently available Noisy Intermediate-Scale Quantum (NISQ) devices. This encoding allows us to trade depth for width in the construction, that is, to use a lesser amount of layers in the preparation circuit if more qubits are available. Due to the depth reduction, the effect of quantum noise can be reduced. In essence, trading depth for width serves as a form of quantum error mitigation in our problem. To verify our claims, we have conducted experiments on the ion-based quantum computers from Quantinuum - the 56-qubit H2-1, H2-2 \cite{Moses2023, DeCross2025} and the latest 98-qubit Helios-1 \cite{ransford2025helios98qubittrappedionquantum}. The obtained results are described in Section \ref{sec:quantinuum} of the paper. 

As a potential genuinely quantum application of RoPE encoding, we consider a DNA authentication problem. In this problem, there are two parties -- a prover and a verifier. The verifier knows a specific DNA sequence, which is either known (up to a certain error rate, say 10\%) or not known by the prover. The prover's task is to convince the verifier that they know the DNA sequence by sending only a single message. The verifier's task is to correctly confirm or reject the prover's message, possibly with a small error probability. What is the optimal way to achieve this using the smallest possible message? In Section \ref{sec:dna-auth}, we estimate the number of qubits required by sending the Angular RoPE encoding of the DNA sequence as proof of knowledge and conjecture that it achieves an exponential reduction in communication complexity over any classical solution. Note that a quantum advantage in one-way communication complexity has been recently proven and demonstrated in \cite{Kretschmer2025}, although for a different communication problem.

Finally, we remark that our experiments with the Angular encoding can serve as a form of verification and benchmarking of a quantum computer itself. Although the sizes of the executed quantum circuits do not allow for exact classical simulation, the results of the execution are somewhat predictable by construction (as we seek the correlation between fidelities and LD). A significant deviation from the predicted outputs could indicate an issue, while matching results signal that the quantum computer is working correctly. 
\section{RoPE-based encodings of DNA sequences}\label{sec:rope}

Rotary Position Embeddings (RoPE) \cite{Su2024RoFormer} are the modern version of positional encoders (PE) that add positional information to text embeddings for use in the transformer architecture, originally introduced in the foundational paper \cite{Vaswani2017Attention}. Their basic idea can be described as follows: 
assume we have a text with tokens from some vocabulary, and a context window is selected. 
The tokens have initial vector embeddings $x_i \in \mathbb{R}^d$ that do not depend 
neither on the token position $i$ in the context window, nor on their neighbors tokens.  
For a specifically constructed rotary matrix $R$, acting on the embedding space $\mathbb{R}^d$, 
we can consider rotated embeddings $y_i = R^i x_i$ which do possess positional information. 
What is important is that $R$ has angles which are all close to $0$, yet different. 
In the original proposal they are defined by the formula $\theta_i = 10000^{-2i / d}, i = 0, \dots, d/2 - 1$.  
This implies that the inner product $y_j^T y_i = x_j^T R^{i-j} x_i$ is approximately the same as $x_j^T x_i$
if positions $i,j$ are close, but keeps drifting away if the relative distance increases. 
The actual use of such rotated embeddings in transformer's attention mechanism is more involved and out of scope of this paper. 

In this section, we describe our main result --- a RoPE-motivated encoding of DNA sequences 
that provides a strong correlation property between the Levenshtein distance and fidelity between encodings. 
A more direct comparison to the original RoPE could be found in Subsection \ref{sec:rope-llm}.  

\subsection{Main feature vector}

Suppose that we have a DNA sequence $\textbf{dna}$ of length $N$. 
Let $s$ denote the length of a small sub-kmer (say, $s=5$). 
There are $N-s+1$ possible locations for it in the 
sequence $\textbf{dna}$. We index them by the numbers 
from the set $[N-s+1] = \{0, \dots, N-s\}$.

Consider an arbitrary $s$-mer $\textbf{mer}$ (say, $\textbf{mer} = {\rm ACCGT}$) 
and look for all of its locations 
$\{\textbf{loc}_1, \dots, \textbf{loc}_{n_{\textbf{mer}}}\} \in [N-s+1]$ within $\textbf{dna}$, 
where $n_{\textbf{mer}}$ is the total number of appearances of 
$\textbf{mer}$. Next, compute the following complex number:

\begin{equation}
\begin{split}
    c_{\textbf{mer}} = \sum_{j=1}^{n_{\textbf{mer}}} \exp\left(\frac{2\pi i \cdot \textbf{loc}_j}{N}\right) = \sum_{j=1}^{n_{\textbf{mer}}} \omega^{\textbf{loc}_j}, \\
    \text{where } \omega=\exp\left(\frac{2\pi i}{N}\right).
\end{split}
\end{equation}

It is clear that this number does not change much if the 
locations $\textbf{loc}_j$ drift a bit. Also, it will not change 
significantly if a small number of 
locations disappear and some are introduced. 
Thus, this number is a feature that encodes all locations 
of an $s$-mer and which is robust against small perturbations.

\begin{figure}[h!]
    \centering
    \includegraphics[width=0.8\linewidth]{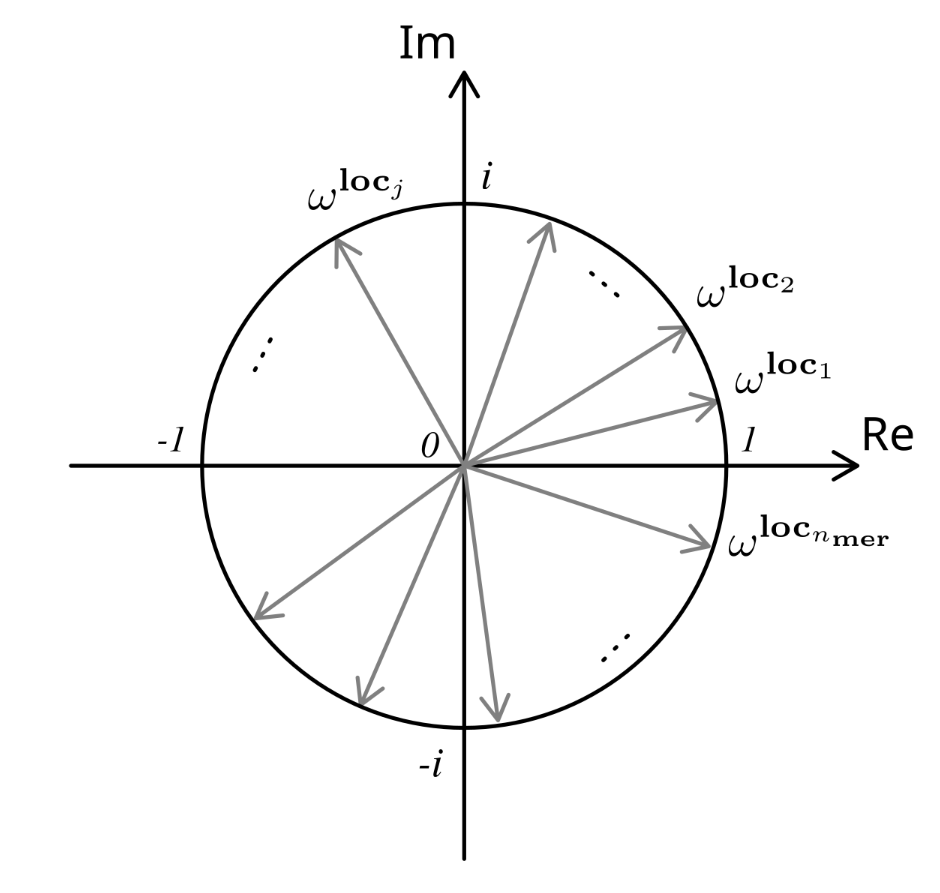}
    \caption{
$c_{\textbf{mer}}$ is a sum of complex units that correspond to locations of $\textbf{mer}$ within the $\textbf{dna}$ sequence.
    }
    \label{fig:rope:circle}
\end{figure}
    
This feature is a basic building block of our encoding. 
We can consider all possible $s$-mers (there are $4^s$ of them), 
compute $c_{\textbf{mer}}$ for each and concatenate them, 
forming a single complex feature vector ${\bf c}$ of size $4^s$.

\subsection{Multiplicity factor}

The defined feature vector may not be sufficient 
to distinguish different $\textbf{dna}$ sequences of the same size. 
We can add additional information in the following way. 
Let $k$ be a small number. Define:
\begin{equation}\label{eq:mult-factor}
c_{\textbf{mer}}^{(k)} = \sum_{j=1}^{n_{\textbf{mer}}} \omega ^{k \cdot \textbf{loc}_j}.
\end{equation}

In essence, we are stretching the set of locations $\textbf{loc}_j$ 
by $k$ times and mapping them back onto the circle \ref{fig:rope:circle}. 
Clearly, numbers $c_{\textbf{mer}}^{(k)}$ are robust against 
perturbations of $\textbf{dna}$ just as well, if $k$ is not too large. However, they differ 
for different factors $k$ and thus contain more information 
if combined together. For a factor $k$, we define the feature 
vector ${\bf c}^{(k)}$ as the concatenation of $c_{\textbf{mer}}^{(k)}$ 
for each possible $s$-mer $\textbf{mer}$.

\subsection{Default version}

\begin{figure}[h!]
    \centering
    \includegraphics[width=1\linewidth]{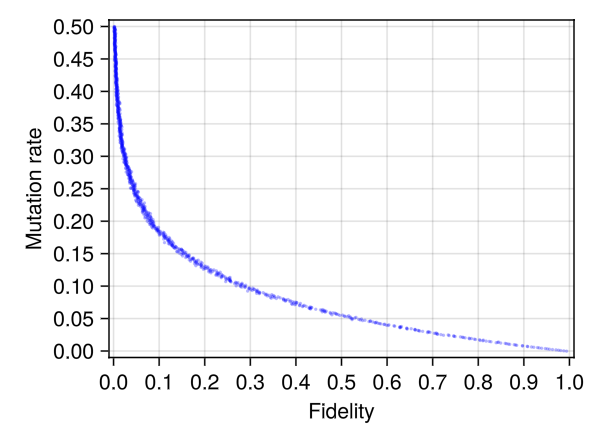}
    \caption{For 1000 random 20,000-long DNA sequences and their mutations, we plot the mutation rate on one axis and fidelity between the default version of RoPE encodings with parameters $s=7, m=4$ (65536 complex dimensions or 16 qubits) on the other. 
    }
    \label{fig:rope_default}
\end{figure}

For parameters $s$ and $m$ (the maximal multiplicity 
factor) we define the default version of the RoPE encoding of 
$\textbf{dna}$ by concatenating feature vectors ${\bf c}^{(k)}$ for all 
$k = 1, \dots, m$ and normalizing the resulting complex 
vector. The dimension of the result is thus 
$m \cdot 4^s$. If $m=2^q$ is a power of $2$, 
the resulting complex vector can be realized as a quantum state 
on $2s+q$ qubits. The observed properties of such an encoding, 
discussed below, lead us to call it the quantum fingerprint 
of a DNA sequence. We denote this state by 
$|{\rm RoPE}_{s,m}(\textbf{dna})\rangle$. 
For example, Figure \ref{fig:rope_default} shows the obtained correlation when the default version of RoPE on 16 qubits is used if DNA strings are random. We discuss the case of non-random DNA strings in Section \ref{sec:rotormap}. 

\begin{figure}[h!]
    \centering
    \includegraphics[width=1\linewidth]{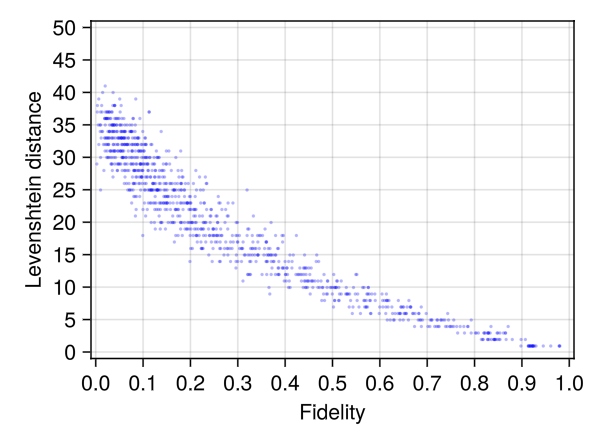}
    \caption{On this plot, random sequences of length 100 bp are encoded with the default RoPE with parameters $s=4,m=4$.
    }
    \label{fig:rope_short}
\end{figure}

The RoPE encoding can also work for short sequences (e.g., of length 100 bp), although the correlation is worse in general, see Figure \ref{fig:rope_short}. 

\subsection{Scoring the correlations}

Note that in our random tests, DNA strings are random (each letter is selected independently with the uniform distribution), while the introduced mutations in their companions are evenly split between random deletions, insertions and substitutions, where the mutation rate is selected evenly from the interval (0, 0.5).

\begin{figure}[h!]
    \centering
    \includegraphics[width=1\linewidth]{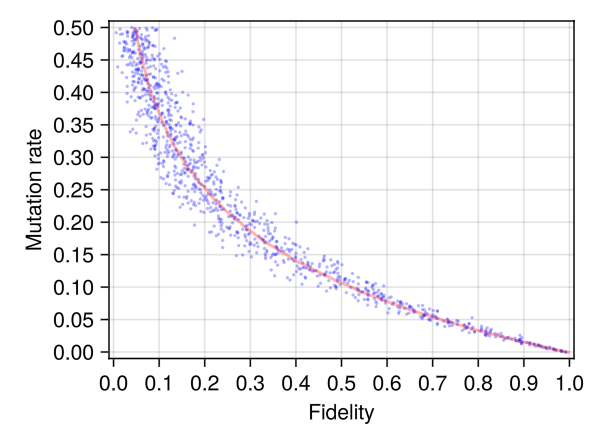}
    \caption{The fitting curve (red) to the correlation obtained from RoPEs with parameters $s=4,m=4$ encoding 1000 bp long random DNA sequences.
    }
    \label{fig:fitting_curve}
\end{figure}

To quantitatively measure obtained correlations, we employ the following score:

\begin{enumerate}
\item For a fixed DNA length $N$, mutation rate $\textbf{err}$ and the 
version of RoPE, we estimate the expectation $\mathbb{E}_\text{fid}(\textbf{err})$ of 
the fidelity $\textbf{fid} = |\langle \text{RoPE}(\textbf{dna})|\text{RoPE}(\textbf{mut}) \rangle|^2 $ when 
sampling pairs $(\textbf{dna}, \textbf{mut})$ randomly.
\item For values $\textbf{err}$ selected evenly from the interval 
$(0, 0.5)$, we compute 2D points $(\mathbb{E}_\text{fid}(\textbf{err}), \textbf{err})$ and connect 
them forming the best fitting curve to the correlation \ref{fig:fitting_curve}. 
\item For a value $\textbf{err}$ of the mutation rate we sample pairs $(\textbf{dna}, \textbf{mut})$, compute 
points $(\textbf{fid}, \textbf{err})$, find vertical distances to the fitting curve and then square them. 
The square root of the average of these values (RMSE) gives an estimate of the error we get when trying to predict the 
mutation rate from the observed fidelity. This is how Figure \ref{fig:intro_pred} was obtained. 
\end{enumerate}

We report the following key observations:

\begin{itemize}
\item The obtained correlation naturally depends on the 
resulting RoPE dimension --- the higher the better. If the output dimensions match, the encoding with a higher $m$ is better, although not by much. Considering that it takes more time to compute, an encoding with a higher $m$ may not always be advantageous. 
\item The smaller the mutation rate, the better the correlation 
for a fixed encoding. In other words, the prediction error decreases if the mutation rate decreases. 
\item The increase in the DNA length $N$ does not affect the 
correlation much relative to $N$ (unless $N$ is too small). This means that 
by looking at the fidelity for a pair $(\textbf{dna}, \textbf{mut})$ one can 
predict the mutation rate in $\textbf{mut}$ with about the same accuracy relative to $N$. 
Perhaps surprisingly, this is true even for $N$ on the order of billions, as Figure \ref{fig:intro_1bil} illustrates. 
\end{itemize}

\subsection{Shift-invariance}

A careful reader might notice that if $m=1$ then the feature 
vector ${\bf c}$ as a function of $\textbf{dna}$ is approximately 
cyclically shift-invariant up to a global phase. 
That is, if we cyclically shift $\textbf{dna}$ by $l$ positions to the 
right and denote the result by $\textbf{dna}_l$, then:
\begin{equation}
  {\bf c}(\textbf{dna}_l) \approx {\bf c}(\textbf{dna}) \cdot \omega^l.
\end{equation}

This means that for the fidelity we have 
$|\langle {\rm RoPE}_{s,1}(\textbf{dna}_l)|{\rm RoPE}_{s,1}(\textbf{dna}) \rangle|^2 \approx 1$. 
One could modify the definition of the RoPE encoding to be 
exactly shift-invariant, though we have not found a meaningful 
benefit to this. 

However, when $m > 1$ the shift-invariance 
breaks. This is because 
${\bf c}^{(k)}(\textbf{dna}_l) \approx {\bf c}^{(k)}(\textbf{dna}) \cdot \omega^{k l}$, 
i.e., feature vectors for different factors $k$ are scaled by 
different phases. And so the fidelity between 
$|{\rm RoPE}_{s,m}(\textbf{dna}_l) \rangle$ and $|{\rm RoPE}_{s,m}(\textbf{dna}) \rangle$ 
will not be close to $1$.

One could try to fix this issue by somehow canonicalizing 
the phase of each ${\bf c}^{(k)}(\textbf{dna}_l)$; however, we have not found 
a way of doing it without breaking the correlation property 
for dissimilar DNA sequences. Thus, a reasonable way to retain 
shift-invariance when $m>1$ is to simply not concatenate 
feature vectors ${\bf c}^{(k)}(\textbf{dna})$, but keep them separate and 
normalized. In such a case, comparing encodings of two 
different DNA sequences means computing fidelities between normalized 
feature vectors ${\bf c}^{(k)}$ for each $k$ and taking the average. 
The approximate shift-invariance may or may not be a desired 
property in applications. 

\subsection{Compact and combined versions}

When the parameter $s$ grows, the dimension of RoPE encoding 
grows exponentially. This may not be a big issue in the quantum 
context, but for classical applications, this is not desired. 
For $s=10, m=1$, the output dimension is on the order of 
$1,000,000$, which is already excessive.

A simple way to treat it is to consider only selected $s$-mers 
when forming feature vectors ${\bf c}^{(k)}$, not all of them. 
For example, for a given reference sequence we could find the 
most frequent $s$-mers in it and use only them in the definition 
of ${\bf c}^{(k)}$. This compact context-aware 
version of the RoPE encoding may be more useful than the 
default version with a smaller $s$. 
Alternatively, for $t < s$ one can use a $4^s$-to-$4^t$ mapping in the algorithm 
to reduce the RoPE output dimension. One could find a link to the reference implementation of this in Section \ref{sec:code}. 

The concatenation of different versions of the RoPE encoding
for different values of parameters $s$ and $m$ could also be 
useful in applications.

\subsection{Strings with uncertainties and other strings}

It is not hard to see that the RoPE encoding could be defined 
for other types of strings just alike, if the alphabet has a 
small size. It may be less convenient to present it as a 
quantum state on many-qubit systems (e.g., if the alphabet size 
is not divisible by $2$), but the selection of $s$-mers as in 
the compact version should solve this problem easily.

In practice, DNA strings can have uncertainties representing 
our missing knowledge. For example, by the IUPAC standard \cite{CornishBowden1985}, 
letter N denotes any of the four main letters A, C, G, T, 
while R represents G or A. We can modify our RoPE encodings 
to represent the semantics of such letters.

Suppose that we see an $s$-mer ANCCG at some location $\textbf{loc}$ 
in the $\textbf{dna}$. A priori, it is reasonable to assume that this 
$s$-mer equals either AACCG, ACCCG, AGCCG, or ATCCG, 
with equal probability. We thus can think that each of these 
$s$-mers appears at the location $\textbf{loc}$ but with the probability 
of $1/4$. And so we modify the definition of $c_{\textbf{mer}}^{(k)}$ 
by setting:
\begin{equation}
  c_{\textbf{mer}}^{(k)} = \sum_{j=1}^{n_{\textbf{mer}}} p(\textbf{mer}, \textbf{loc}_j) \cdot \omega ^{k \cdot \textbf{loc}_j},
\end{equation}
where $p(\textbf{mer}, \textbf{loc})$ is the probability of appearance of $\textbf{mer}$ 
at $\textbf{loc}$.

The same treatment can be applied to base calling quality 
scores (used in the FASTQ format, for example). 
In essence, they modify the probability distribution of the 
observed letter in each position, which we can encode similarly 
to what is explained above. In practice, it could make sense to 
adjust those probabilities to match biological priors.

\subsection{Complexity of computing RoPE encodings}

The computational complexity of generating RoPE encodings is linear with respect to the $\textbf{dna}$ sequence length $N$, as we can slide 
through all possible locations of $s$-mers in a single pass. 
The multiplicity parameter $m$ adds the factor of $m$ to the 
complexity, which is not an issue for small $m$. 
Furthermore, the process is highly parallelizable; the $\textbf{dna}$ sequence can be partitioned into segments for independent processing and subsequent recombination. A batched job that computes encodings for multiple sequences of the same size at the same time fits modern GPU hardware nicely, as branching is fixed and predictable in the algorithm. 

\subsection{Comparing DNA sequences of different lengths}

Given the construction of RoPE encodings, it is clear that directly comparing sequences of different sizes is not meaningful. While one might attempt to modify the encoding to address this, we think it is preferable to employ separate methods that utilize RoPEs as building blocks. Our mapper, which aligns a DNA read to a reference sequence, represents one such method. Similarly, if a semantic analysis of DNA sequences is required, a transformer architecture can be applied on top of the RoPE encodings.

\subsection{Comparison with the original RoPE in LLMs}\label{sec:rope-llm}

Let ${\rm onehot}(\textbf{mer})$ denote the $4^s$-dimensional one-hot 
encoding of an $s$-mer $\textbf{mer}$, ${\rm ones}(k)$ is a vector of ones 
of size $k$, and 
$D = \text{Diagonal}((\omega^1, \dots, \omega^m) \otimes \text{ones}(4^s))$ 
is a diagonal matrix of size $m \cdot 4^s \times m \cdot 4^s$ 
with the specified diagonal. An alternative way to construct 
the default version of RoPE with parameters $s$ and $m$ is 
by taking the sum:
\begin{equation}
  \sum_{\textbf{loc} =0}^{N-s} D^{\textbf{loc}} \cdot \text{ones}(m) \otimes \text{onehot}(\text{mer}(\textbf{loc})),
\end{equation}
where $\text{mer}(\textbf{loc})$ is the $\textbf{mer}$ at $\textbf{loc}$, and normalizing in 
the end.

Essentially, our tokens consist of sequences of $s$ consecutive letters, which are initially encoded as stretched one-hot vectors and subsequently rotated by the unitary diagonal matrix $D^{\textbf{loc}}$ depending 
on their location. While one could attempt to define a variant of RoPE that aligns more closely with the original formulation by altering token definitions, initial encodings, or the rotation matrix, we have not identified a combination superior to the one presented in this work.

\section{RotorMap --- a RoPE-based DNA mapper}\label{sec:rotormap}

Given a reference DNA sequence and a DNA read 
(a smaller fragment sequenced from biological material), 
the mapping problem objective is to find an approximate matching 
position of the read in the reference sequence -- if one exists. 
Here we present a novel classical algorithm, named RotorMap, 
that solves this problem using the RoPE-DNA encodings.

\subsection{Basic idea}

First, let us fix a length $N$ that is shorter than the typical length of the DNA reads we expect to map. Imagine a window of size $N$ sliding across the reference sequence $\textbf{ref}$ with a defined step size $\textbf{step}$, selecting a set of $N$-mers that cover the reference sequence. For each of these $N$-mers, we compute the RoPE encoding, creating an index (essentially a database of vectors). For a given DNA read, we extract its initial segment (head) of length $N$, compute its RoPE encoding, then compute fidelities between the encoding and all the vectors within the index -- which is just a matrix multiplication at its core. Fidelities larger than a certain threshold value allow us to determine matching approximate locations. 

While one could employ modern vector indexing techniques, such as those based on the popular HNSW algorithm~\cite{Malkov2020}, we have found that our proof-of-concept (PoC) implementation, which is equivalent to a linear database scan, is already fast and produces good results, especially when mapping long reads with at least 20,000 base pairs. 

\subsection{Performance estimation}

Since the computation of RoPE encodinds is linear in the DNA length and has a simple branching its speed is negligible in comparison to the search phase. For example, the encoding with parameters $s=4, m=4$ of a 1 million reads of length 20,000 bp each should take on the order of 80 billion arithmetic operations times some constant. In our PoC implentation it takes less than 2 seconds on a single H100 NVIDIA GPU and probably could be optimized further.

Let us consider the mapping of DNA reads of length 20,000 bp to a reference sequence of size 3.3 billion. A window of size $N=20,000$ sliding with a step size of 1,250 (6.25\% of the read length) will select approximately 2,600,000 $N$-mers. Consequently, computing RoPE encodings with an output dimension of 1024 yields a database containing approximately 2,600,000 1024-dimensional complex vectors.

Given the RoPE encoding of a single read, a simple linear scan through the entire database to find the best match thus requires about 10 billion arithmetic operations. Hence, a batch of 100,000 reads should be mapped in about 1 second on a device capable of 1 PetaFLOPS performance. Since the core part of the search is simply a matrix product, this speed is within the reach on modern GPU and TPU devices. In our PoC implementation, such a search phase (with some additional postprocessing) takes less than 20 seconds on a single H100 NVIDIA GPU using FP16 precision.

Mapping shorter reads is a more challenging problem. Assume that we have to map $t$ times shorter reads. If the window sliding step is also $t$ times shorter (which is reasonable), then the total number of selected N-mers will be $t$ times larger. And thus, the linear scan through this database will take $t$ times longer (apart from requiring $t$ times more memory to store the index). Therefore, the use of a vector database is a more suitable solution in this case.

\subsection{Dealing with periodic DNA}

\begin{figure}[h!]
    \centering
    \includegraphics[width=1\linewidth]{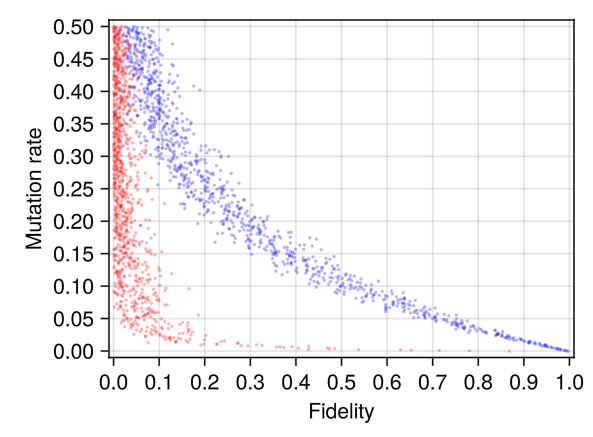}
    \caption{The correlation colored red is obtained for mutations of a 20,000 long fragment related to the D4Z4 macrosatellite repeats. Default RoPE-DNA with parameters $s=4,m=4$ is used. The blue correlation is obtained for random DNA strings. 
    }
    \label{fig:rope_nonrandom_bad}
\end{figure}

While the RoPE encoding works especially good for random DNA sequences, it is not the case in general. The sum of evenly spaced complex units on the unit circle is zero, thus the default version of RoPE encoding of the sequence with only $A$ letters is the zero vector, making it useless. A periodic DNA fragment with the period that divides the fragment length would produce the zero vector as well. Even if a sequence is not exactly periodic with the matching period, it results in a kind of singularity. For example, in tests it was found that a 20,000-long fragment in chomosome 4 of the human genome has a period of approximately 3,300. It consists of D4Z4 macrosatellite repeats related to the DUX4 gene. If we fix this DNA fragment and consider its random mutations with various mutation rates, the correlation provided by the default version of RoPE could be seen in Figure \ref{fig:rope_nonrandom_bad}. It is apparent that this correlation is not very useful in practice. The human genome has many other places with periodic fragments, most famously telomeres and centromeres. 

One might think that a set of random complex units on the unit circle, if evenly distributed, should have the same problem as in the periodic scenario. This is not the case, however. The sum of $n$ independent random complex units (which have angles uniformly distributed on $[0, 2\pi)$) can be approximated by a complex normal distribution with mean 0 and variance $n$ as $n$ grows large, due to the Central Limit Theorem. The modulus of such a complex random variable follows the Rayleigh distribution with scale parameter $\sqrt{n/2}$ and has mean value of $\sqrt{n\pi}/2$. 

One could try various methods of dealing with the periodicity issue. A simple solution that works sufficiently well in practice is the following. Instead of Equation \ref{eq:mult-factor}, defining the default version of RoPE, we can consider its fine-tuned version: 
\begin{equation}\label{eq:mult-factor-finetune}
c_{\textbf{mer}}^{(k)} = \sum_{j=1}^{n_{\textbf{mer}}} \omega ^{(2(k-1)/(m-1)+1) \cdot \textbf{loc}_j}.
\end{equation}

\begin{figure}[h!]
    \centering
    \includegraphics[width=1\linewidth]{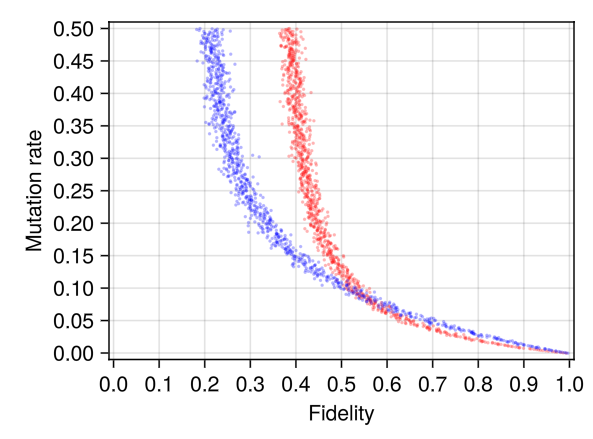}
    \caption{The correlation colored red is obtained for mutations of the same 20,000-long fragment related to D4Z4, as shown in Figure \ref{fig:rope_nonrandom_bad}. A compact, fine-tuned version of RoPE-DNA with parameters $s=8, m=4, t=4$ is used. The blue correlation is obtained for random DNA strings.
    }
    \label{fig:rope_nonrandom_good}
\end{figure}

This fix essentially stretches the set of complex units that corresponds to s-mer locations by different fractional factors. 
As a result, the values $c_{\textbf{mer}}^{(k)}$ become biased, but are not simply related to each other, as in the case of a periodic sequence. 
Figure \ref{fig:rope_nonrandom_good} shows how the new correlations look.

\begin{figure}[h!]
    \centering
    \includegraphics[width=1\linewidth]{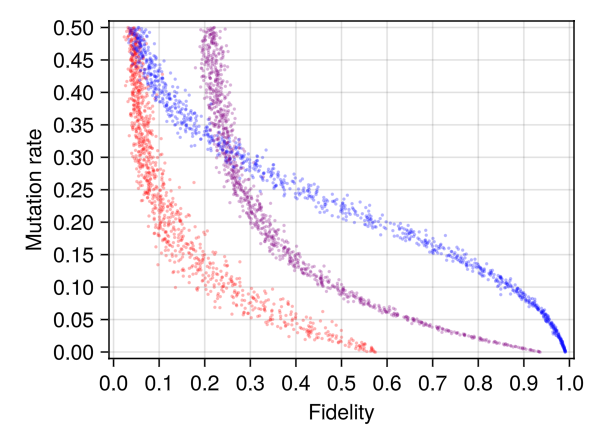}
    \caption{The performance of fine-tuned RoPE-DNA with parameters $s=8,m=4,t=4$. In each of the three correlations, random mutations are applied after an additional {\it shift} mutation by 500 bp. The purple correlation is obtained for random DNA strings, while the red and blue correlations correspond to the lowest and highest fidelities related to a 0.10 mutation rate, among 20,000-long DNA fragments covering the human genome. 
    }
    \label{fig:rope_finetuned}
\end{figure}

In a comprehensive test, we checked the correlations provided by the fine-tuned RoPE-DNA for 20,000-long fragments covering the entire human genome. 
Among them, we identified the two extreme correlations with the lowest and highest fidelities that correspond to a 0.10 mutation rate, see Figure \ref{fig:rope_finetuned}. In this figure, random mutations are applied after an additional random {\it shift} mutation, which simply cuts the head and adds a random tail of the same length to the sequence. Any other correlation should fall somewhere between these two extremes.

Note that the different forms of correlations, as in Figure \ref{fig:rope_finetuned}, are not an issue in a mapping algorithm. For a particular target mutation rate (which defines whether there is similarity or not), we can estimate the corresponding values of fidelity for each fragment in the reference genome and use them as cutting thresholds when deciding on similarity based on fidelity.

\subsection{Increasing the accuracy}

If we use a step of 10\% of the read length in the construction of the index, then we can expect to have less than 5\% difference between found and true location (that provides the best match in a region), assuming the RoPE encoding have sufficient number of dimensions. 

A simple way to increase accuracy is to recompute the index around found locations using a smaller step, say 0.1\% of the read length, and then perform a comparison with the read encoding again. A straightforward way of doing it would not be too costly, however, another remarkable property of our RoPE encoding allow us to do it significantly faster.  

Assume for a moment that we use $m=1$ in the RoPE encoding and that ${\bf x} = (x_{\textbf{mer}})$ is the $4^s$-dimensional main feature vector of the DNA read, while $X$ is its squared norm. Similarly, define ${\bf c} = (c_{\textbf{mer}})$ to be the main feature vector of the reference fragment at found approximate location, with $C$ being its squared norm. 

Let us consider what happens when we slide the window, that selects the found fragment from the reference, by 1 bp to the right. The position of each $s$-mer $\textbf{mer}$ in the window gets decreased by 1, except the first one which leaves the scope. Also, a new $s$-mer enters the window from the very right end. Using the definition of RoPE, we get the following relations:

\begin{align}
c_{\textbf{mer}}^{new} = c_{\textbf{mer}}\omega^{-1}, 
\end{align}
for each {\bf mer} except two 
\begin{align}
c_{\textbf{oldmer@start}}^{new} = c_{\textbf{oldmer@start}}\omega^{-1} - \omega^{-1}, \\
c_{\textbf{newmer@end}}^{new} = c_{\textbf{newmer@end}}\omega^{-1} + \omega^{N-s},
\end{align}
or in the case if two $s$-mer exceptions match
\begin{align}
c_{\textbf{mer-matched}}^{new} = c_{\textbf{mer-matched}}\omega^{-1}  - \omega^{-1} + \omega^{N-s}.
\end{align}

The inner product between ${\bf c}$ and ${\bf x}$ is defined by $\text{ip} = \sum_{\textbf{mer}} \overline{x}_\textbf{mer} c_\textbf{mer}$ and thus changes by the relation 
\begin{align}
\text{ip}^{new} 
= \text{ip}\omega^{-1} 
- \overline{x}_{\textbf{oldmer@start}}\omega^{-1} 
+ \overline{x}_{\textbf{newmer@end}}\omega^{N-s}.
\end{align}

The new fidelity is then simply 
\begin{align}
\text{fid}^{new} = \frac{1}{XC^{new}}|\text{ip}^{new}|^2.
\end{align}

The computation of the new squared norm $C^{new}$ of $\textbf{c}^{new}$ requires keeping the track of changes of each $c_{\textbf{mer}}$. This could be done lazily as we know by how much they are rotated at each slide step. In practice, however, norm recomputation can be skipped since it does not change much if we are already close enough to a good match.

For a different multiplicity factor $k$, everything is basically the same, except we should use $\omega^k$ instead of $\omega$ in the default RoPE or $\omega^{(2(k-1)/(m-1)+1)}$ in the fine-tuned. 

So, it turns out that the recomputation of the fidelity between encodings when we slide through the reference sequence has a small constant cost that does not really depend neither on the read length nor on the RoPE dimension (except for the factor $m$). This allows for quickly finding a significantly more accurate mapped location using a high-dimensional RoPE, e.g., with 65536 complex dimensions. 

\subsection{Comparison with Minimap2}

To estimate the potential speedup of RotorMap over Minimap2 in a real scenario, we conducted tests against the human reference genome assembly GRCh38.p14 (GenBank accession: GCA\_000001405.29) and the Zea mays B73 reference genome assembly Zm-B73-REFERENCE-NAM-5.0 (RefSeq accession: GCF\_902167145.1), retrieved from the NCBI database \cite{NCBI2015}. The sets of test reads covering each genome were generated using the PBSIM software \cite{Ono2022}, with a target read length of 24 kbp (subsequently cut to 20 kbp) and an error rate of 15\% using the ONT profile. They contain approximately 96,000 simulated reads from the human genome and approximately 64,000 from the maize genome. The hardware used was part of the Wellcome Sanger Institute HPC infrastructure. Table \ref{tab:speed_comparison} shows a comparison of approximate average running times for different settings. 

\begin{table}[h]
\small
\centering
\caption{Mapping Runtime Comparison (Seconds)}
\label{tab:narrow_mapping}
\begin{tabular}{@{}l|c|r@{}}
\toprule
\textbf{Method} & \textbf{Hardware/Threads} & \textbf{Time (s)} \\ \midrule
\multicolumn{3}{l}{\textbf{Human Genome} (96k reads, 20kbp each)} \\ \midrule
RotorMap & 1$\times$ H100 GPU & $< 40$ \\
Minimap2 & 80 CPU Threads & $\approx 50$ \\
Minimap2 & 1 CPU Thread & $\approx 2,000$ \\ \midrule
\multicolumn{3}{l}{\textbf{Maize Genome} (64k reads, 20kbp each)} \\ \midrule
RotorMap & 1$\times$ H100 GPU & $< 20$ \\
Minimap2 & 80 CPU Threads & $\approx 280$ \\
Minimap2 & 1 CPU Thread & $\approx 14,000$ \\ \bottomrule
\end{tabular}
\label{tab:speed_comparison}
\end{table}

\begin{table*}[t]
\centering
\begin{tabular}{|l|c|r|r|r|r|}
\hline
\textbf{Genome} & \textbf{Error Rate} & \textbf{Number of Reads} & \textbf{True Matches} & \textbf{Other Matches} & \textbf{Mismatches} \\ \hline
Random & 20\% & 100,000 & 100,000 (100\%) & 0 (0\%)& 0 (0\%)\\ \hline
Human & 15\% & 182,566 & 154,479 ($\approx$ 85\%) & 12,547 ($\approx$ 7\%)& 15,540 ($\approx$ 8\%) \\ \hline
Human & 10\% & 181,924 & 164,319 ($\approx$ 90\%) & 12,688 ($\approx$ 7\%)& 4,917 ($\approx$ 3\%)\\ \hline
Human & 5\% & 180,852 & 166,677 ($\approx$ 92\%) & 12,674 ($\approx$ 7\%)& 1,501 ($\approx$ 1\%)\\ \hline
Maize & 15\% & 127,539 & 113,677 ($\approx$ 89\%) & 2,920 ($\approx$ 2\%)&  10,942 ($\approx$ 9\%)\\ \hline
Maize & 10\% & 127,139 & 120,634 ($\approx$ 95\%) & 2,977 ($\approx$ 2\%)& 3,528 ($\approx$ 3\%)\\ \hline
Maize & 5\% & 126,593 & 122,671 ($\approx$ 97\%) & 2,865 ($\approx$ 2\%)& 1,057 ($\approx$ 1\%)\\ \hline
\end{tabular}
\caption{Accuracy of the mapping: if the mapped location overlaps with the true location, it is a true match; if not, but the Levenshtein distance is low, it is another good match; otherwise, it is a mismatch.}
\label{table:rotormap-accuracy}
\end{table*}

To test the mapping accuracy, sets of reads were generated with varying error rates (see Table \ref{table:rotormap-accuracy}). In these tests, RotorMap operated in a regime that output only a single location corresponding to the highest fidelity between a read and the index (without using the accuracy amplification stage). If the returned location overlapped with the true location from which the read was sampled, it was considered a true match. If the location was different, yet the Levenshtein distance between the read and the found fragment was not high ($<30$\%), it was still counted as a match. Otherwise, it was considered a mismatch. Note that in tests against a randomly generated reference sequence (with a uniform distribution of letters) of length 1 billion, the results were perfect, yielding 100\% of true matches. The fine-tuned RoPE with parameters $s=8, m=4, t=4$ (1024 complex dimensions) was used in the algorithm. The size of sliding step for index construction was set to $20,000/16=1250$. 

RotorMap can also operate in a regime where it outputs all possible matching locations based on the fidelity threshold values, precomputed for each fragment in the index. In this regime, the number of missed matches can be reduced because of an improved probability that the found locations contain the true location for each read. However, this results in an increased number of false positives in the results, which must be filtered. For example, in a test of mapping a batch of human reads with a 15\% error rate (the same as in Table \ref{table:rotormap-accuracy}), conducted with fidelity threshold values estimated specifically for this error rate, we saw an average of about 6.6 found locations per read, although about 78\% of reads were mapped to exactly one location, which was almost surely correct. However, the output contained more than a thousand found locations in some rare cases, most of which were false positives. In this regime, the fraction of missed matches (reads for which the found locations do not contain the true location) was about 6\%. This value could be improved with better estimation of fidelity threshold values.
\section{Angular encoding -- a method to prepare RoPEs}\label{sec:angular}

\begin{figure*}
    \includegraphics[width=1\linewidth]{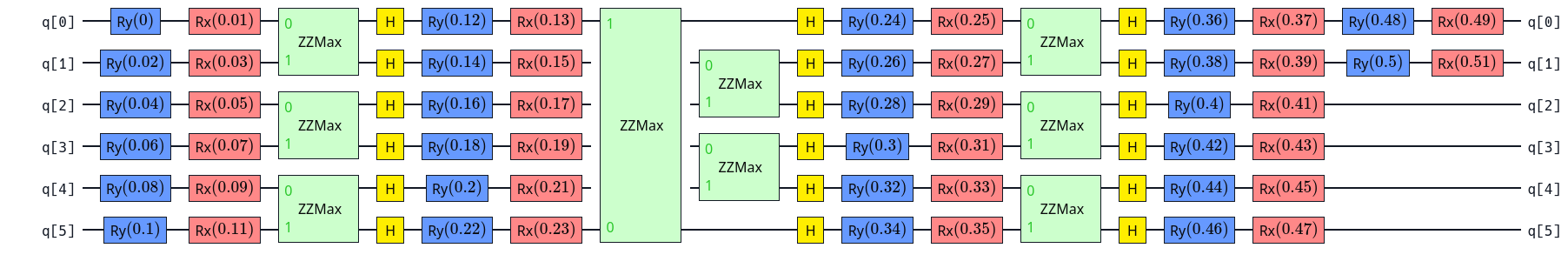}
    \caption{
    The standard variation of Angular encoding. The example is constructed from a complex vector of dimension 26 (it does not have to be a power of 2), targeting 6 qubits. The parameters indicate the order; in reality, they correspond to the real and imaginary parts of the complex numbers from the vector.
    }
    \label{fig:angular-design}
\end{figure*}

Since the RoPE-DNA encoding constitutes a normalized complex vector, it can be treated as a quantum state. However, to utilize this state on quantum devices, a preparation method is required. That is, we need to find a quantum circuit $U$ such that $U|\bf{0}\rangle = |{\rm RoPE}(\textbf{dna}) \rangle$. Identifying such a circuit is known as the state preparation problem. Because the RoPE encoding lacks symmetries, we should treat it as an arbitrary state. In general, an exponential number of gates is required to prepare such states exactly, see \cite{Mottonen2004}.

Approximate state preparation is a practical alternative to exact preparation. One of the promising directions in solving this problem is based on using Matrix Product States (MPS), see, for example, \cite{creevey2025scalablequantumstatepreparation} and references therein. Even though approximate state preparation methods can noticeably reduce the size of the constructed circuits, the scaling is still exponential, and they are also computationally heavy.

Luckily, in our DNA encoding problem, a different kind of optimization is possible. Since our primary concern is the correlation between edit distance and fidelity, we do not have to prepare RoPE states as they are; any state that retains the correlation property serves our purpose equally well. This insight forms the basis of our Angular encoding, which is built from RoPE but outputs different, transformed states.

First, note that there is a natural isomorphism between complex vectors in $\mathbb{C}^d$ and real vectors in $\mathbb{R}^{2d}$. For a vector $|z\rangle = |x\rangle + i|y\rangle \in \mathbb{C}^d$, $|x\rangle, |y\rangle \in \mathbb{R}^{d}$, its image under this isomorphism is given by $|w\rangle = \text{Re}(|z\rangle) \oplus \text{Im}(|z\rangle) = |x\rangle \oplus |y\rangle \in \mathbb{R}^{2d}$. If $|z\rangle$ is normalized, then so is $|w\rangle$. Moreover, for two complex vectors $|z_1\rangle$, $|z_2\rangle$ we have 
\begin{equation}
\begin{split}
  |\langle z_1|z_2 \rangle|^2 &= (\langle x_1|x_2 \rangle + \langle y_1|y_2 \rangle)^2 
  + (\langle x_1|y_2 \rangle - \langle y_1|x_2 \rangle)^2 \\ 
  &\approx |\langle w_1|w_2 \rangle|^2,
\end{split}
\end{equation}
where $(\langle x_1|y_2 \rangle - \langle y_1|x_2 \rangle)^2 \approx 0$ should hold in general because the real and imaginary parts of RoPE encodings (essentially the vectors of cosines and sines) are not similar to each other for random DNA strings. The expectation of fidelity between two independent Haar random vectors in $\mathbb{R}^{d}$ is $\frac{1}{d}$, thus we can write
\begin{equation}
\begin{split}
  & \mathbb{E} (\langle x_1|y_2 \rangle - \langle y_1|x_2 \rangle)^2 \le \\
  & \mathbb{E} |\langle x_1|y_2 \rangle|^2 + 2\mathbb{E}|\langle x_1|y_2 \rangle||\langle y_1|x_2 \rangle| + \mathbb{E}|\langle y_1|x_2 \rangle|^2 \\
  & = \frac{4}{d},
\end{split}
\end{equation}
assuming $|x_1 \rangle,|x_2 \rangle,|y_1 \rangle,|y_2 \rangle$ are Haar random and $|x_i \rangle$ are independent of $|y_j \rangle$. 
Edge cases with a disproportional amount of different nucleotide bases, such as sequence with only $A$ in it, may require additional patching and fine-tuning. 

The basic idea of Angular encoding is to take the $2d$-dimensional real vector, extracted from $|{\rm RoPE}(\textbf{dna}) \rangle$, and insert its elements in place of the angles of single-qubit gates in a specifically constructed quantum circuit. The standard variation of the circuit is shown in Figure \ref{fig:angular-design}. It consists of a brickwork pattern where layers of two-qubit gates (specifically, $\text{ZZMax}=e^{-i\frac{\pi}{4}Z\otimes Z}$ gates, as we target Quantinuum devices) are interleaved with layers of single-qubit gates. The single-qubit layers combine Rx, Ry, and H gates, as depicted. In the two-qubit layers, the pairing between qubits is shifted by 1 as we move to the next layer.

The circuit structure was motivated by the observation that to encode a vector of $n$ presumably independent parameters, we must have the same number $n$ of places to insert them in a circuit, otherwise we lose degrees of freedom. This puts a constraint on the number of single-qubit gates in a circuit, which allows for only single-qubit gates and fixed two-qubit gates, such as CNOT or ZZMax. The next question is then how many two-qubit gates we need and how to arrange them. If we do not use any two-qubit gates, then single-qubit gates acting on the same qubit will blend into a single operation, which reduces the freedom of the resulting state as well. The brickwork pattern provides probably the simplest way to prevent blending of single-qubit gates. A trial-and-error methodology was used to find the exact circuit structure that provides a good correlation as a result.

One can notice that the number of gates scales exponentially with the RoPE dimension. However, we are not bounded by the number of qubits in the Angular encoding; we can use any number, in theory. In particular, if we increase the target number of qubits, the depth will decrease roughly by the same factor. This depth-width tradeoff is particularly useful on currently available NISQ devices, as the depth of a circuit is the main factor that contributes to noise, not the total number of gates.

\begin{figure}[h!]
    \centering
    \includegraphics[width=1\linewidth]{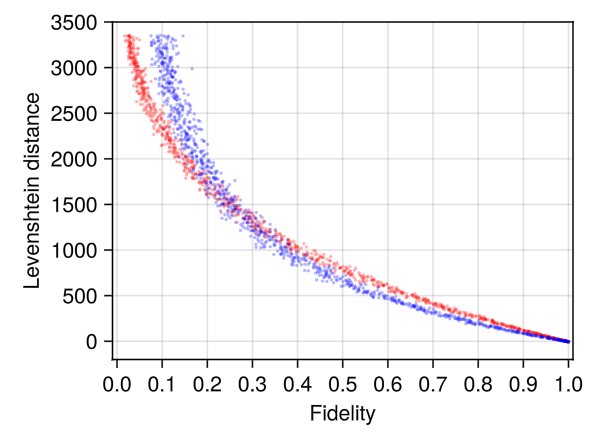}
    \caption{A RoPE encoding of 10,000-mers on 11 qubits (red points) is transformed into the Angular encoding on 20 qubits (blue points).
    }
    \label{fig:angular-compare}
\end{figure}

For example, consider a RoPE encoding on 9 qubits. It has 512 complex numbers, so we need 1024 places for angles in the Angular encoding. If we target 20 qubits, then $\lfloor 1024/2/20 \rfloor = 25$ single-qubit layers are needed (we do not separate the last incomplete layer from the rest) and 24 two-qubit layers. However, the Angular encoding on 56 qubits requires only $\lfloor 1024/2/56 \rfloor = 9$ single-qubit and $8$ two-qubit layers in this case. 
Figure \ref{fig:angular-compare} 
shows noiseless simulation results comparing Angular encoding with RoPE. As can be seen, the correlation still exists, although it is slightly different. Note that before supplying parameters from a RoPE encoding to the Angular encoding, we can apply a scaling factor to them. This factor controls the slope in the resulting correlation, which can be useful in certain tasks.

\subsection{Compact version}

One can consider more compact versions of the Angular encoding. An arbitrary single-qubit gate can be parametrized by three real angles via the decomposition $\text{Rz}(\alpha)\text{Rx}(\beta)\text{Rz}(\gamma)$, 
which gives us three places for parameters. Additionally, instead of the fixed ZZMax gate, one can use the Rxxyyzz gate defined by 
\begin{equation}
    \text{Rxxyyzz}(\alpha, \beta, \gamma)=e^{-i\frac{\pi}{2}(\alpha X\otimes X + \beta Y\otimes Y + \gamma Z\otimes Z)}, 
\end{equation}
 which also has three places for parameters. This gate is called TK2 on Quantinuum H2 systems; it is executed with additional guaranties that prevent unnecessary ion transport. 

\begin{figure}[h!]
    \centering
    \includegraphics[width=1\linewidth]{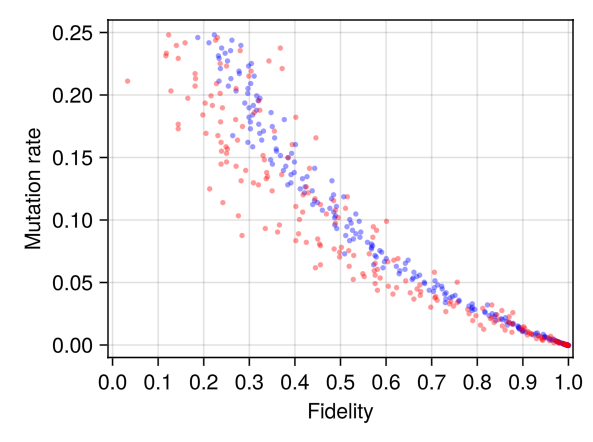}
    \caption{Standard Angular encoding (blue) versus a compact version (red), noiseless simulation. Both target 20 qubits and were constructed from the same 9-qubit RoPE encodings of random 20,000-mer pairs.
    }
    \label{fig:angular_compact}
\end{figure}

The resulting correlation, provided by compact versions of Angular encoding, does not allow for a good approximation of LD based on the observed fidelity, in general. However, its use could be sufficient if only a possible similarity between DNA fragments is in question, see Figure \ref{fig:angular_compact}, for example.

\section{Results of experiments}\label{sec:quantinuum}

To verify practicality of the introduced DNA encodings on currently available Noisy Intermediate-Scale Quantum (NISQ) devices, we have conducted experiments on Quantinuum's H2 and Helios systems (\href{https://www.quantinuum.com/}{https://www.quantinuum.com}), which operate on 56 and 98 qubits, respectively. 

The experiments consist of the following steps:

\begin{enumerate} 
    \item 16 pairs of random 100,000-long DNA strings were sampled, with the mutation rate in pairs selected evenly from the interval (0, 0.25).
    \item The default version of RoPE encoding on 9 qubits with parameters $s=4, m=2$ was computed for the sampled DNA sequences.
    \item The Angular encodings, which represent quantum circuits and target different numbers of qubits in different tests, were constructed from the RoPEs.
    \item To check the fidelity on quantum devices, we applied a circuit corresponding to one DNA in a pair, followed by the conjugate circuit corresponding to the other DNA in the pair, to the zero state $|\bf 0\rangle$.
    \item After conducting multiple shots, the return probability of measuring the zero state was recorded. This probability equals the fidelity between the two encodings exactly. 
\end{enumerate}

One can see a similarity between our experiments and the mirror benchmarking \cite{mayer2023theorymirrorbenchmarkingdemonstration}, a popular technique that measures the performance of quantum computers. There are a couple of noticeable differences:

\begin{itemize}
    \item In our experiments, the pairing between qubits in two-qubit layers is fixed, while mirror benchmarking randomizes it.

    \item Even though the second half of executed circuits has a mirrored structure, the parameters of its single-qubit gates are different (as they come from the RoPE of mutated DNA in a pair). If a compact version of the Angular design is used, the parameters of two-qubit TK2 gates also differ in the mirrored part.

    \item The parameters of gates in our circuits are quite small, as they are elements of a normalized high-dimensional vector. In mirror benchmarks, single-qubit gates are selected as Haar random, so they do not have this restriction.
\end{itemize} 

Additionally, mirror benchmarks usually apply randomized compiling \cite{Wallman2016} to circuits, which we skipped due to high costs of treating each shot separately.

\begin{figure}[h!]
    \centering
    \includegraphics[width=1\linewidth]{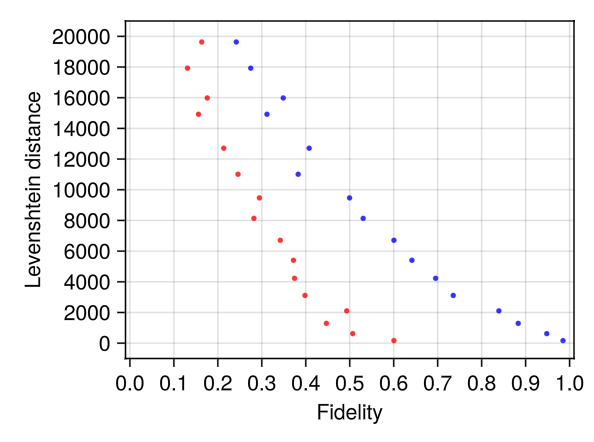}
    \caption{Tests of the standard Angular encoding targeting 28 qubits in noiseless simulation (blue) versus targeting 56 qubits, 800 shots execution on Quantinuum's H2-2 (red).
    }
    \label{fig:angular_main_results}
\end{figure}

\begin{figure}[h!]
    \centering
    \includegraphics[width=1\linewidth]{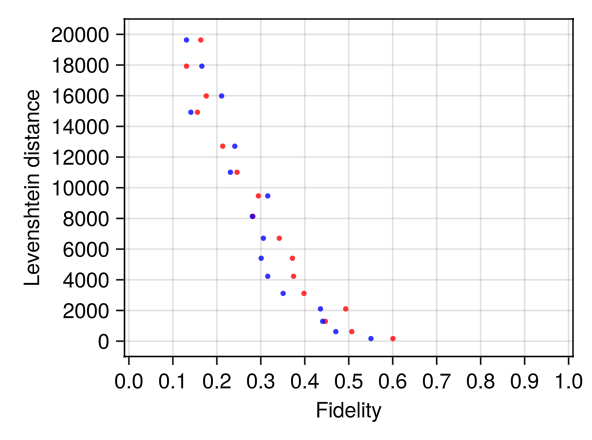}
    \caption{The standard Angular encoding, real execution on 98 qubits on Quantinuum's Helios-1, 200 shots per circuit (blue) versus the standard Angular encoding on 56 qubits on H2-2 (red) --- the same as in Figure \ref{fig:angular_main_results}.
    }
    \label{fig:angular_helios}
\end{figure}

Figures \ref{fig:angular_main_results} and \ref{fig:angular_helios} show results of tests of the described experiments in cases where
\begin{enumerate}
    \item the standard Angular encoding targets 28 qubits, the results were simulated without any noise, which approximates the ground truth;
    \item the target number of qubits is 56, the results were executed on Quantinuum's H2-2 using 800 shots for each circuit on October 21-28, 2025;
    
    \item the target number of qubits is 98, the results were executed on Quantinuum's Helios-1 using 200 shots for each circuit on Mar 12-17, 2026.
\end{enumerate}

The correlation pattern is evident in each of the cases. The average drop in fidelity for real execution in Figure \ref{fig:angular_main_results} is approximately 0.56, which is due to the accumulated quantum errors. Other tests, in particular those targeting 20 qubits on both the H2 and Helios systems and 56 qubits on Helios, generated similar correlation patterns, up to uncertainties. We noticed a slight decrease in the fidelity drop in later tests, as the Quantinuum systems keep improving. 

\begin{figure}[h!]
    \centering
    \includegraphics[width=1\linewidth]{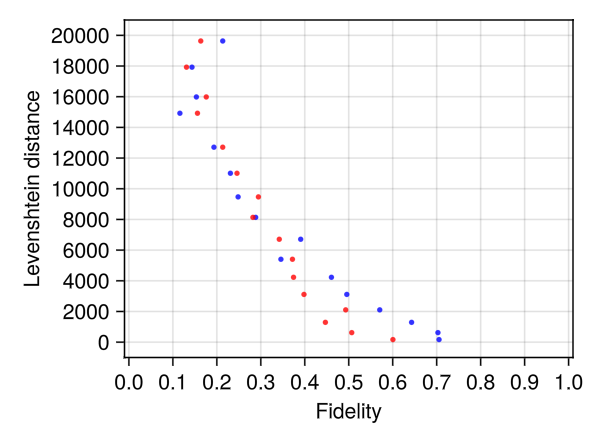}
    \caption{A compact version of Angular encoding, real execution on 56 qubits on Quantinuum's H2-2, 400 shots per circuit (blue) versus the standard version (red), which is the same as in Figure \ref{fig:angular_main_results}.
    }
    \label{fig:angular_compact_real}
\end{figure}

We have also tested a compact version of the Angular encoding on 56 qubits on Quantinuum's H2-1, using 400 shots per circuit on February 9-13, 2026. The reduced depth (6 instead of 16) and the use of TK2 gates instead of ZZMax caused its results to appear slightly more advantageous than those of the standard Angular encoding, despite its worse performance in noiseless simulations. As can be seen in Figure \ref{fig:angular_compact_real}, points with lower Levenshtein distances have slightly higher fidelity values for the compact Angular encoding than for the standard, which should allow for better separation from points with high Levenshtein distances.

Note that while the Angular encoding is capable of the width-depth trade-off, there is no advantage in using wider circuits on current Quantinuum devices. This is largely because the number of gate zones is relatively small - four on H2 and eight on Helios - allowing for the processing of, respectively, eight and sixteen qubits simultaneously. However, the planned future devices from Quantinuum have more gate zones and thus should allow for the parallel processing of many more qubits. This is where we expect to have an advantage in using wider circuits in the Angular encoding.
\section{DNA authentication problem}\label{sec:dna-auth} 

As mentioned in the introduction, there are two parties in the DNA authentication problem: a prover and a verifier. The prover must convince the verifier that they know the same DNA sequence up to some Levenshtein distance by sending \textit{a single message}. In turn, the verifier must correctly confirm or reject it, possibly with some small error probability. What is the smallest message that can accomplish this?

To enable theoretical analysis of this problem, we must make some adjustments. First, it makes sense to formulate it as a gap edit distance problem, where the verifier has to decide if the Levenshtein distance $d$ between their DNA and the prover's DNA is either below some threshold $a$ (say, $10\%$ of the sequence length $N$) or larger than some value $b$ (say, $30\%$ of $N$). The verifier is not required to answer correctly if $a \leq d \leq b$. Secondly, we assume that the verifier's DNA is generated randomly by selecting each letter uniformly and independently.

A naive way to solve this problem classically is by sending the entire sequence, which has a communication complexity of $2N$ bits. The optimal complexity for arbitrary values of $a$ and $b$ is not known. However, for the gap Hamming distance problem between bit strings with $a = N/2 - \sqrt{N}$ and $b = N/2 + \sqrt{N}$, the complexity is $\Omega(N)$, even if we allow multiple rounds of communication (and do not impose any restrictions on the strings), see \cite{Sherstov2012}. In this case, there is a known quantum advantage, as a Grover-like quantum algorithm was constructed in \cite{Buhrman1998} that solves it in $O(\sqrt{N} \log(N))$. 

\begin{figure}[h!]
    \centering
    \includegraphics[width=0.8\linewidth]{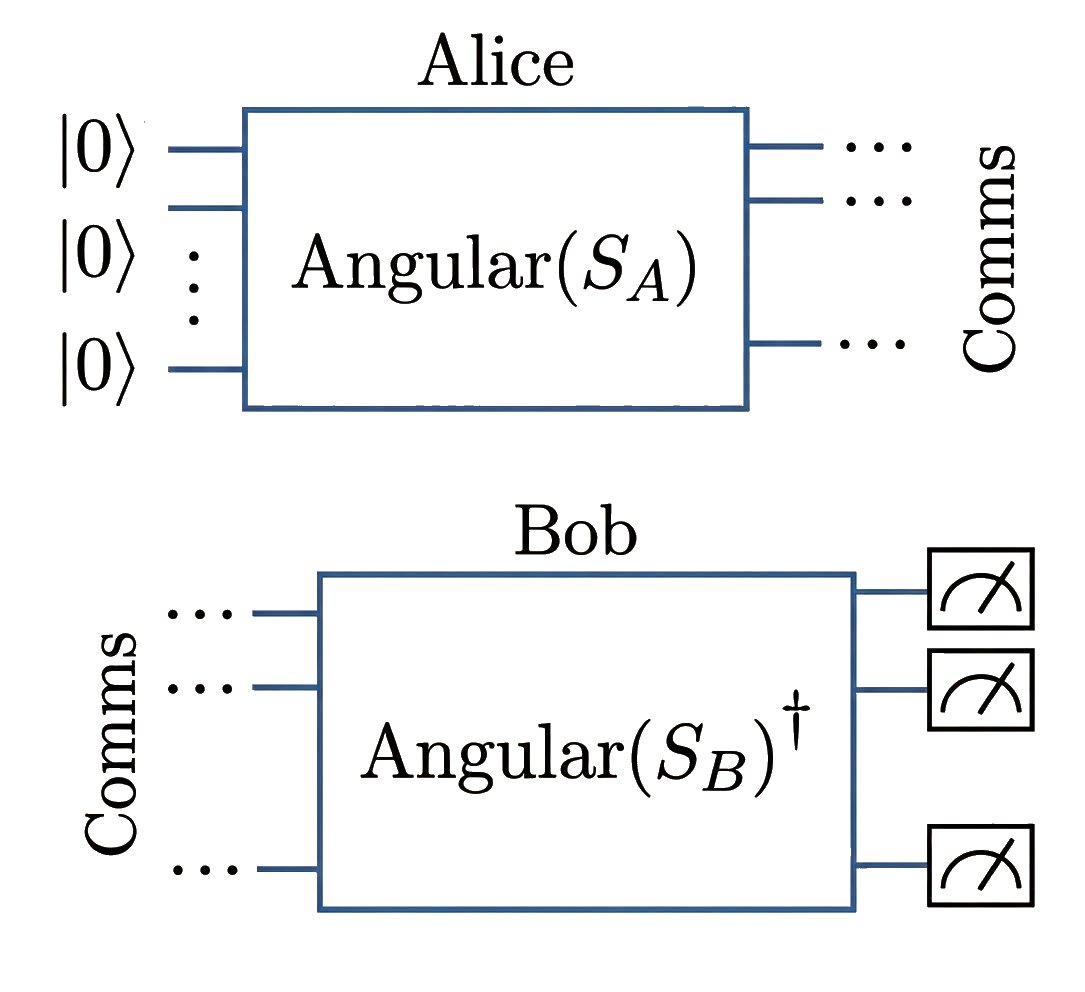}
    \caption{A scheme for quantum DNA authentication: Alice (the prover) encodes her sequence $S_A$ using Angular encoding and sends it to Bob, the verifier. Upon receiving it, Bob applies the conjugate encoding of his sequence $S_B$ and measures the result, noticing the probability of obtaining the zero state.
    }
    \label{fig:dna-auth-scheme}
\end{figure}

With access to quantum devices, the prover can send multiple copies of the RoPE/Angular quantum encoding of their DNA. The verifier simply has to apply the inverse preparation circuit for their DNA to the received states and measure the result, noting the probability of obtaining the zero state, see Figure \ref{fig:dna-auth-scheme}. This probability is equal to the fidelity between the encodings, and so the verification can be decided based on it. This verification protocol is essentially what was done in our experiments, as described in Section \ref{sec:quantinuum}, assuming that both the prover and the verifier have access to the same device and that communication occurs immediately.

We argue that this achieves a very low quantum communication complexity. Let us assume that $N=10^9$, $a=0.1N$, and $b=0.3N$. As one can see in Figures \ref{fig:intro_corr} and \ref{fig:intro_1bil}, the correlation patterns for $N=20,000$ and $N=10^9$ are similar. We thus assume that the correlation in the latter case has similar characteristics (in particular, the prediction error) to that in the former. The prediction error \ref{fig:intro_pred} is less than $1.5\%$, which suggests that 12 qubits in the RoPE encoding is well enough to separate the error rates of $10\%$ and $30\%$ based on fidelity. For $N=10^9$, the corresponding fidelities are about $0.45$ and $0.1$, respectively. It is natural to assume that if the fidelity is $f$, then in each measurement (each shot), the observation of the zero state is modeled by the Bernoulli distribution with mean $f$. By using Hoeffding's inequality, one could find that to distinguish between two Bernoulli random variables with means $f_1$ and $f_2$, it is enough to make $\frac{-2\ln\varepsilon}{(\Delta f)^2}$ samples, where $\Delta f = |f_1-f_2|$ and $\varepsilon$ is the error probability. So, $\frac{-2\ln0.01}{0.35^2} \approx 75$ shots are enough to be correct $99\%$ of the time, which corresponds to a total communication complexity of $12 \cdot 75 = 900$ qubits. 

In our experiments on Quantinuum, described in Section \ref{sec:quantinuum}, we have observed a drop of approximately 0.56 in fidelity due to quantum noise when using standard Angular encoding on 20 qubits (in comparison to the noiseless Angular 28-qubit case, which is close to the 12-qubit RoPE). Thus, we require approximately $1/0.56^2 \approx 3.2$ more shots to solve the problem. This amounts to a total of $20 \cdot 75 \cdot 3.2 = 4,800$ qubits of communication complexity on Quantinuum's H2 systems.

In the general case, we conjecture that the one-way quantum communication complexity in the DNA authentication problem with randomly sampled strings does not depend on $N$ but rather on the relative gap $(b-a)/N$. Note that the prover could send a classical description of the RoPE encoding instead of the quantum state. This requires $2^{12}\cdot2\cdot16 = 131,072$ classical bits if the FP16 precision is used for the real and imaginary parts of the complex entries in the encoding. This might be the optimal way to solve this problem classically. In this case, the quantum advantage that we have conjectured is exponential in magnitude.

\section{Discussion} \label{sec:discussion}

The RoPE-based encoding of DNA sequences, which is our main contribution in this paper, opens new ways of DNA information processing in general and on quantum computers in particular. While we report mostly experimental findings, we believe that theoretical guarantees can be proven for our or similar RoPE-motivated constructions. It would be interesting to discover what makes RoPE so effective and improve upon that. Perhaps an LD approximation algorithm with a guaranteed linear time could be constructed.

The practicality of RoPE-DNA was demonstrated with RotorMap, a new GPU-accelerated DNA mapping algorithm that outperforms Minimap2 in our proof-of-concept implementation when mapping reads of length 20,000 bp to human and maize genomes by roughly two and three orders of magnitude, respectively. While the increase in speed should be much less dramatic (if any) for short reads of length 100 bp, RotorMap has great potential for improvement if combined with a vector database for storage and search of RoPE encodings.

The RoPE encoding of DNA fragments can also be used in DNA language models that extract semantic information from DNA sequences. Unlike texts in natural languages, DNA sequences lack delimiters and punctuation, which leads to difficulties in the direct application of large language model techniques. See \cite{Testagrose2025} for a recent review of tokenization methods in deep learning in genomics. In particular, it seems that RoPE-DNA could fit architectures that use overlapping k-mers for tokenization, except that $k$ could be much longer than usual.

For use on quantum devices, we have presented the Angular encoding, which, given a RoPE encoding, outputs a preparation circuit (albeit for a different, transformed state). Our tests on Quantinuum hardware show that correlation between LD and fidelity between encodings can be observed despite present quantum noise in the system. This opens up possibilities for using RoPE-DNA in quantum algorithms and protocols, as in the proposed DNA authentication problem, for which we conjecture that there is a quantum advantage in the RoPE-based solution.
\section{Acknowledgments}\label{sec:acknowledgements}

This work was supported by Wellcome Leap as part of the Q4Bio Program. The authors thank James MacCafferty for support throughout the project, as well as Richard Durbin, James Bonfield, Robert Davies, and the rest of the QPG team for insightful discussions. We are also grateful to Alina Frolova for support and to the Quantinuum team for their help and assistance with quantum experiments. 
S.S. was additionally supported by the Royal Society University Research Fellowship.



\section{Data and code availability}\label{sec:code}
A Julia function that computes RoPE encodings with parameters $s,m,t$ of DNA sequences on a CPU can be found at \href{https://github.com/dandanua/rope-dna/}{github.com/dandanua/rope-dna/}. 
Experimental data could be provided upon reasonable request.

\bibliography{RoPE.bib}


\appendix*

\end{document}